\documentclass{jfm}
\usepackage{graphicx, color}
\usepackage{epstopdf, epsfig}
\usepackage{amsmath}

\usepackage{floatrow}
\usepackage{subfig}
\usepackage{multirow}
\usepackage{longtable}

\usepackage{cleveref}
\crefname{section}{section}{sections}
\crefname{subsection}{subsection}{subsections}
\crefname{figure}{figure}{figures}
\crefname{table}{table}{tables}
\crefname{equation}{}{}
\Crefname{section}{Section}{Sections}
\Crefname{subsection}{Subsection}{Subsections}
\Crefname{figure}{Figure}{Figures}
\Crefname{table}{Table}{Tables}

\definecolor{mygreen}{rgb}{0,.5,0}
\definecolor{mygray}{rgb}{.5,.5,.5}
\definecolor{mybrown}{rgb}{.43,.21,.1}

\renewcommand{\thetable}{\arabic{table}S}

\newcommand{\beq}{\begin{equation}}
\newcommand{\eeq}{\end{equation}}

\newcommand{\uu}{\mathbf{u}}
\newcommand{\cO}{\mathcal{O}}
\newcommand{\rmd}{\mathrm{d}}



\newcommand\solidrule[1][15pt]{\rule[0.5ex]{#1}{1pt}}
\newcommand\dashedrule{\mbox{%
	\solidrule[3pt]\hspace{3pt}\solidrule[3pt]\hspace{3pt}\solidrule[3pt]}}

\newcommand\dottedrule{\mbox{%
	\solidrule[1.5pt]\hspace{2pt}\solidrule[1.5pt]\hspace{2pt}\solidrule[1.5pt]\hspace{2pt}\solidrule[1.5pt]\hspace{2pt}\solidrule[1.5pt]}}

\shorttitle{Steady Rayleigh--B\'enard convection between no-slip boundaries}
\shortauthor{B. Wen, D. Goluskin and C. R. Doering}

\title{Steady Rayleigh--B\'enard convection between no-slip boundaries}

\author{
\mbox{Baole Wen}\aff{1}\corresp{\email{baolew@umich.edu}},
\mbox{David Goluskin}\aff{2}\corresp{\email{goluskin@uvic.ca}},
 \and
\mbox{Charles R.\ Doering}\aff{1,3,4}}

\affiliation{
\aff{1}Department of Mathematics, University of Michigan, Ann Arbor, MI 48109-1043, USA
\aff{2}Department of Mathematics \& Statistics, University of Victoria, Victoria, BC, V8P 5C2, Canada
\aff{3}Department of Physics, University of Michigan, Ann Arbor, MI 48109-1040, USA
\aff{4}Center for the Study of Complex Systems, University of Michigan, Ann Arbor, MI 48109-1042, USA}

\begin{document}

\maketitle

\begin{abstract}
The central open question about Rayleigh--B\'enard convection---buoyancy-driven flow in a fluid layer heated from below and cooled from above---is how vertical heat flux depends on the imposed temperature gradient in the strongly nonlinear regime where the flows are typically turbulent. The quantitative challenge is to determine how the Nusselt number $Nu$ depends on the Rayleigh number $Ra$ in the $Ra\to\infty$ limit for fluids of fixed finite Prandtl number $\Pran$ in fixed spatial domains. Laboratory experiments, numerical simulations, and analysis of Rayleigh's mathematical model have yet to rule out either of the proposed `classical' $Nu \sim Ra^{1/3}$ or `ultimate' $Nu \sim Ra^{1/2}$ asymptotic scaling theories. Among the many solutions of the equations of motion at high $Ra$ are steady convection rolls that are dynamically unstable but share features of the turbulent attractor. We have computed these steady solutions for $Ra$ up to $10^{14}$ with $\Pran=1$ and various horizontal periods. By choosing the horizontal period of these rolls at each $Ra$ to maximize $Nu$, we find that steady convection rolls achieve classical asymptotic scaling. Moreover, they transport more heat than turbulent convection in experiments or simulations at comparable parameters. If heat transport in turbulent convection continues to be dominated by heat transport in steady rolls as $Ra\to\infty$, it cannot achieve the ultimate scaling.
\end{abstract}

\vspace{-0.4in}

\section{Introduction}
Rayleigh--B\'enard convection (RBC) is the buoyancy-driven flow in a fluid layer heated from below and cooled from above in the presence of gravity. The emergent convective flow enhances heat flux from the warm bottom boundary to the cool top boundary beyond the conductive flux from diffusion alone. This dimensionless enhancement factor---the ratio of bulk-averaged vertical heat flux from both conduction and convection to the flux from conduction alone---defines the Nusselt number $Nu$. In Rayleigh's mathematical model \citep{Rayleigh1916} $Nu$ depends on several dimensionless quantities characterizing the problem at hand: (i) what we now call the Rayleigh number $Ra$, which is proportional to the imposed temperature drop across the layer, (ii) the fluid's Prandtl number $Pr$, which is the ratio of kinematic viscosity to thermal diffusivity, and (iii) details of the spatial domain, often captured by an aspect ratio $\Gamma$ that is a ratio of a horizontal length scale to the vertical layer height.

Convection is {\it coherent} at $Ra$ values not too far above the critical value $Ra_c$ beyond which the conductive no-flow state is linearly unstable. By coherent we mean flows with few scales present; spatial scales might include a horizontal period and the vertical thickness of boundary layers, and temporally the flow may be steady or time-periodic. Meanwhile, convection is {\it turbulent} at the large $Ra$ values pertinent to many engineering and scientific applications. Turbulent flows are complex and contain a range of spatial and temporal scales and, in the present context, have thermal and viscous boundary layers at the top and bottom boundaries from which thermal plumes emerge and mix the bulk. In a given domain it is expected that a scaling of $Nu$ with respect to both $Pr$ and $Ra$ will emerge in the $Ra\to\infty$ limit \citep{Kadanoff2001}.

After nearly a century of increasingly sophisticated mathematical analysis, increasingly resolved direct numerical simulations (DNS), and increasingly refined laboratory experiments, two quantitatively distinct conjectures remain in contention for the heat transport scaling law at large $Ra$ \citep{Chilla2012, Doering2020}. The two conjectures follow from heuristic physical arguments that both seem plausible but give incompatible predictions: the `classical' scaling $Nu \sim Pr^0 Ra^{1/3}$ and the `ultimate' scaling $Nu \sim Pr^{1/2} Ra^{1/2}$, with the latter sometimes including logarithmic-in-$Ra$ modifications.

For RBC between flat, no-slip, isothermal boundaries, rigorous analysis of the governing equations has yielded upper bounds of the form $Nu\le \cO(Ra^{1/2})$ uniformly in $Pr$ and $\Gamma$ \citep{H63,DC1996}, but this still allows for either classical or ultimate scaling.
Upper bounds that rule out ultimate scaling by being asymptotically smaller than ${\cal O}(Ra^{1/2})$ have been derived in the limit of infinite $\Pran$ \citep{DoeringOttoReznikoff2006, Otto2011a, Whitehead2012} and for two-dimensional convection between stress-free boundaries \citep{Whitehead2011}. For the no-slip boundaries relevant to experiments, however, it remains an open question whether an upper bound asymptotically smaller than $Ra^{1/2}$ is possible.

\begin{figure}[t]
\centering
\includegraphics[width=1\textwidth]{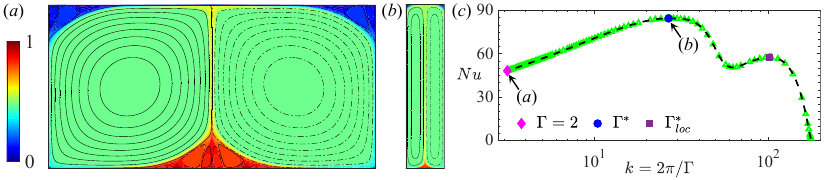}
\vspace{-0.2in}
\caption{\label{fig: rolls}Steady convection rolls at $Ra=10^9$ and $Pr=1$ with ($a$) $\Gamma = 2$ and ($b$) the value $\Gamma^*\approx0.235$ that maximizes $Nu$ at these values of $Ra$ and $Pr$. Color indicates temperature, and streamlines are shown for counterclockwise (solid) and clockwise (dash-dotted) motions. ($c$) Dependence of $Nu$ on the horizontal wavenumber $k=2\pi/\Gamma$ found by computing steady rolls of various aspect ratios (${\scriptstyle\color{green}\triangle}$). Highlighted points are the two $\Gamma$ shown in panels ($a$) and ($b$) along with the value $\Gamma^*_{loc}\approx0.0614$ that locally maximizes $Nu(\Gamma)$. Cubic spline interpolation ($\dashedrule$) is used to find $\Gamma^*$ and $\Gamma^*_{loc}$ precisely.}
\end{figure}

In view of the problem's stubbornness, a new strategy is called for to determine---or at least to bound---$Nu$ as a function of $Ra, Pr$, and $\Gamma$. Toward that end we have undertaken an indirect approach consisting of two parts. The first part is to study coherent flows for which one can reasonably hope to determine asymptotic heat transport, and the second part is to investigate how transport by those coherent flows compares with transport by turbulent convection. The simplest coherent flows are steady---i.e., time-independent---solutions of the equations of motion. Many such states exist, although they are generally unstable at large $Ra$. We focus on what might be called the simplest type of steady states: two-dimensional convection rolls like the counter-rotating pairs shown in \cref{fig: rolls}($a$, $b$). In horizontally periodic or infinite domains in two or three dimensions, such rolls bifurcate supercritically from the conductive state in the linear instability identified by \cite{Rayleigh1916}.  A roll pair of any width-to-height aspect ratio $\Gamma$ admitted by the domain exists for sufficiently large $Ra$.

For steady rolls, the dependence of $Nu$ on the parameters $(\Gamma,Pr, Ra)$ at \emph{asymptotically large} $Ra$ is accessible to computation. As for whether heat transport by steady rolls can be connected to transport by turbulence, there are several reasons for optimism.
Relationships between turbulent attractors and the unstable coherent states embedded therein have been established in models of wall-bounded shear flows \citep{Graham2021}, where particular steady states, traveling waves, and time-periodic states have been found that closely reflect turbulent flows in terms of integral quantities as well as particular flow structures. Analogous study of RBC began only recently but indeed suggests that certain steady states capture qualitative aspects of turbulent convection~\citep{Waleffe2015, Sondak2015, Kooloth2021, Motoki2021}. Our findings add to this evidence. The desire to understand and perhaps strengthen the mathematical bound $Nu\le \cO(Ra^{1/2})$ is further motivation for studying unstable states since bounds apply to all solutions of the governing equations regardless of stability. It is an open question whether \emph{any} solutions can achieve ultimate scaling, let alone turbulent solutions.

Here we report numerical computations of steady convection rolls for a $Pr=1$ fluid contained between no-slip isothermal top and bottom boundaries. We reach sufficiently large $Ra$ values to convincingly reveal several asymptotic scalings of $Nu$, depending on the horizontal periods of the rolls. These are the first clearly asymptotic scalings found for any type of flow---steady, turbulent, or otherwise---for RBC in the no-slip case. Notably, the largest heat transport among steady rolls of all horizontal periods displays the classical $Nu\sim Ra^{1/3}$ scaling. We further observe that $Nu$ for these steady rolls is \emph{larger} than turbulent $Nu$ from all laboratory experiments and two- or three-dimensional (2D or 3D) simulations at comparable parameters. This observation supports the conjecture that steady states maximize $Nu$ among all stable or unstable flows, as was recently verified for a truncated model of RBC \citep{Olson2021} using methods that are not yet applicable to the full governing equations. If steady-roll transport continues to dominate turbulent transport as $Ra\to\infty$, then our finding of classical scaling for steady rolls would rule out ultimate scaling of turbulent convection.

The asymptotic scaling of steady rolls is already known in the case of \emph{stress-free} velocity conditions at the top and bottom boundaries, which were considered for mathematical convenience in Rayleigh's original work. In that case $Nu\sim Ra^{1/3}$ as $Ra\to\infty$ at fixed $Pr$ and $\Gamma$, and the aspect ratio of the roll pair maximizing $Nu$ at each $Ra$ and $Pr$ approaches $\Gamma \approx 1.9$ \citep{ChiniCox2009, Wen2020}.  Recent computations of steady rolls in the {\it no-slip} case for pre-asymptotic $Ra$ values up to $10^9$ revealed significant differences from the stress-free problem \citep{Waleffe2015, Sondak2015}. The dependence $Nu(\Gamma)$ for no-slip rolls at fixed $Ra$ and $Pr$ can have multiple local maxima, as shown in \cref{fig: rolls}($c$), and the aspect ratio $\Gamma^*$ that globally maximizes $Nu(\Gamma)$ approaches zero rather than a constant as $Ra\to\infty$. Steady rolls of $Nu$-maximizing aspect ratios $\Gamma^*$ were reported in \cite{Sondak2015} for $Ra \in [5\times10^6, 3\times10^8]$ at $Pr=1$, yielding fits of $\Gamma^* \sim Ra^{-0.217}$ and $Nu(\Gamma^*) \sim Ra^{0.31}$. This heat transport scaling is faster than with $\Gamma$ fixed: computations in \cite{Waleffe2015} for $Ra \in [5\times10^5, 5\times10^6]$ at $Pr=7$ with $\Gamma = 2$ fixed yield the fit $Nu\sim Ra^{0.28}$.  These best-fit scaling exponents are, however, not asymptotic.

Steady convection rolls are dynamically unstable at large $Ra$ and cannot be found by standard time integration, so we employed a purpose-written code that iteratively solves the time-independent  equations. We computed rolls with $\Gamma=2$ fixed for $Ra\lesssim2\times10^{10}$ and with the parameter-dependent aspect ratios $\Gamma^*$ and $\Gamma^*_{loc}$ (cf.\ \cref{fig: rolls}) that globally and locally maximize $Nu(\Gamma)$, respectively, for $Ra\le10^{14}$. These $Ra$ values are evidently large enough to reach asymptotia: the results reported below strongly suggest that fixed-$\Gamma$ rolls asymptotically transport heat like $Nu\sim Ra^{1/4}$ while the ever-narrowing rolls of aspect ratio $\Gamma^*$ achieve the classical $Nu\sim Ra^{1/3}$ scaling.

\section{Computation of steady-convection-roll solutions}

Following \cite{Rayleigh1916}, we model RBC using the Boussinesq approximation to the Navier--Stokes equations with constant kinematic viscosity $\nu$, thermal diffusivity $\kappa$, and coefficient of thermal expansion $\alpha$. We nondimensionalize lengths by the layer height $h$, temperatures by the fixed difference $\Delta$ between the boundaries, velocities by the free-fall scale $U_f = \sqrt{g\alpha h\Delta}$, and time by the free-fall time $h/U_f$. Calling the horizontal coordinate $x$ and the vertical coordinate $z$, the gravitational acceleration of magnitude $g$ is in the $-\mathbf{\hat z}$ direction. The evolution equations governing the dimensionless velocity vector $\uu = (u, w)$, temperature $T$, and pressure $p$ are then
\begin{subequations}
\begin{align}
\partial_t u + \uu \cdot \nabla \uu  &= 
-\nabla p + ({Pr}/{Ra})^{1/2} \;\nabla^2 \uu + T \mathbf{\hat z}, \label{eq: u} \\
\nabla \cdot \uu &= 0, \label{eq: inc} \\
\partial_t T + \uu \cdot \nabla T& = (Pr Ra)^{-1/2}\; \nabla^2 T, \label{eq: T}
\end{align}
\label{eq: bouss}
\end{subequations}
where
\begin{subequations}
\begin{align}
   Ra = \frac{g\alpha h^3\Delta}{\kappa\nu} \quad \mbox{and}  \quad Pr = \frac{\nu}{\kappa}.	\tag{\theequation $a$,$b$}
\end{align}
\end{subequations}
The dimensionless spatial domain is $(x,z)\in[0,\Gamma]\times[0,1]$, and all variables are horizontally periodic.
The top and bottom boundaries are isothermal with $T=0$ and $T=1$, respectively, while no-slip conditions require $\uu$ to vanish on both boundaries. The conductive state $(\uu,T)=(\mathbf 0,1-z)$ becomes unstable when $Ra$ increases past the critical value $Ra_c\approx 1708$ \citep{Jeffreys1928}, at which a roll pair with horizontal period $\Gamma \approx 2.016$ bifurcates supercritically. As $Ra\to\infty$ the horizontal period of the narrowest marginally stable roll pair decreases as $\cO(Ra^{-1/4})$, while the horizontal period of the fastest-growing linearly unstable mode decreases more slowly as $\cO(Ra^{-1/8})$.

In terms of the dimensionless solutions to \cref{eq: bouss}, the Nusselt number is 
\beq
Nu = 1 + (Pr Ra)^{1/2} \, \langle wT\rangle,
\label{eq: Nu}
\eeq
where $\langle\cdot\rangle$ denotes an average over the spatial domain and infinite time. For steady states no time average is needed.

To compute rolls at $Ra$ values large enough to reach the asymptotic regime we developed a numerical scheme by adapting the approach of \cite{Wen2020} and \cite{WenChini2018JFM} to the case of no-slip boundary conditions. In these numerics the temperature is represented using the deviation $\theta$ from the conductive profile, meaning $T=1-z+\theta$, and the velocity is represented using a stream function $\psi$, where $\uu = \partial_z \psi \mathbf{\hat x} - \partial_x \psi \mathbf{\hat z}$ so that the (negative) scalar vorticity is $\omega=\partial_x w- \partial_z u = -\nabla^2 \psi$. In terms of these variables, steady ($\partial_t=0$) solutions of \cref{eq: bouss} satisfy
\begin{subequations}
\label{eq: steadybouss}
\begin{align}
\partial_z\psi\partial_x\omega - \partial_x\psi\partial_z\omega  &= 
({Pr}/{Ra})^{1/2}\; \nabla^2 \omega + \partial_x \theta, \label{eq: omega} \\
\nabla^2 \psi &= -\omega, \label{eq: psi}\\
\partial_z\psi\partial_x\theta - \partial_x\psi\partial_z\theta & = -\partial_x\psi + (Pr Ra)^{-1/2}\; \nabla^2 \theta \label{eq: theta}
\end{align}
\end{subequations}
with fixed-temperature and no-slip boundary conditions,
\begin{subequations}
\label{eq: BCs}
\begin{align}
\theta|_{z=0, 1}=0,  \qquad \psi|_{z=0,1}=0, \qquad \text{and} \qquad \partial_z \psi|_{z=0,1}=0.		\tag{\theequation $a$,$b$,$c$}
\end{align}
\end{subequations}

To compute solutions of the time-independent equations \cref{eq: steadybouss,eq: BCs} by an iterative method, we do not need to impose all boundary conditions precisely on each iteration---the conditions need to hold only for the converged solution. Thus we do not impose (\ref{eq: BCs}$c$) exactly, instead using approximate boundary conditions on $\omega$ for equation \cref{eq: omega}. These are derived by Taylor expanding $\psi$ about the top and bottom boundaries to find
\begin{subequations}
\label{eq: Psi_Taylor}
\begin{align}
\psi|_{z=1-\delta} &= \psi|_{z=1} - {\partial_z \psi}\big|_{z=1} \delta + {\partial^2_{z} \psi}\big|_{z=1} \frac{\delta^2}{2}  - {\partial^3_z \psi}\big|_{z=1} \frac{\delta^3}{6} + \cO(\delta^4), \label{eq: Psi_top}\\
\psi|_{z=\delta} &= \psi|_{z=0} + {\partial_z \psi}\big|_{z=0} \delta + {\partial^2_z \psi}\big|_{z=0} \frac{\delta^2}{2}  + {\partial^3_z \psi}\big|_{z=0} \frac{\delta^3}{6} + \cO(\delta^4), \label{eq: Psi_bot}
\end{align}
\end{subequations}
where $\delta>0$ is small.  Combining equations~(\ref{eq: psi}) with (\ref{eq: BCs}$b$,$c$) and neglecting $\cO(\delta^4)$ terms in \cref{eq: Psi_Taylor} give the approximate boundary conditions,
\begin{subequations}
\label{eq: Psi_Taylor2}
\begin{align}
\partial_z\omega|_{z=1} - \frac{3}{\delta}\ \omega|_{z=1}  -  \frac{6}{\delta^3}\ \psi|_{z=1-\delta} = 0, \quad
-\partial_z\omega|_{z=0} - \frac{3}{\delta}\ \omega|_{z=0}  -  \frac{6}{\delta^3}\ \psi|_{z=\delta} = 0.  \tag{\theequation $a$,$b$}
\end{align}
\end{subequations}
In computations we set $\delta$ to be the distance between the boundary and the first interior mesh point.

The time-independent equations (\ref{eq: steadybouss}) are solved numerically subject to boundary conditions (\ref{eq: BCs}$a$,$b$) and (\ref{eq: Psi_Taylor2}) using a Newton--GMRES (generalized minimal residual) iterative scheme. The spatial discretization is spectral, using a Fourier series in $x$ and a Chebyshev collocation method in $z$ \citep{Trefethen2000}. All of our computations had at least 20 collocation points in the viscous and thermal boundary layers. At $Ra$ just above the linear instability, iterations starting from the unstable eigenmode converge to the steady rolls we seek. At larger $Ra$, already-computed steady rolls from nearby $Ra$ and $\Gamma$ values were used as the initial iterate. 
Every 2 to 4 Newton iterations, we change the boundary values of the iterate to match the $\partial_z\psi=0$ boundary condition exactly. Prior to convergence this makes the boundary values slightly inconsistent with the governing equations, but the converged solutions satisfy the equations and the no-slip boundary conditions to high precision. Newton iterations were carried out until the Lebesgue $L^2$-norm of the residual of the governing steady equations had a relative magnitude less than $10^{-10}$. To accurately locate $\Gamma^*$ and $\Gamma^*_{loc}$, rolls were computed at several nearby $\Gamma$, and then $Nu(\Gamma)$ was interpolated with cubic splines like those in \cref{fig: rolls}($c$). Details of computational results, including resolutions used, are included in the supplementary material.

\section{Results}

We computed steady $Pr=1$ rolls for aspect ratios $\Gamma$ encompassing the three distinguished values indicated by \cref{fig: rolls}($c$): the fixed value $\Gamma=2$ and the $Ra$-dependent values $\Gamma^*$ and $\Gamma^*_{loc}$ that globally and locally maximize $Nu$ over $\Gamma$. As previously observed by \cite{Sondak2015}, the $Nu(\Gamma)$ curve has a single maximum when $Ra$ is small and develops a second local maximum at smaller $\Gamma$ when $Ra$ increases past roughly $2\times10^5$. The value of $Nu$ at this second local maximum remains less than the value at the first, so the picture remains as in \cref{fig: rolls}($c$) with $\Gamma^*$ on the left and $\Gamma^*_{loc}$ on the right, in contrast to the $Pr=10$ and 100 cases \citep{Sondak2015}. For most $Ra$ values we did not compute rolls over a full sweep through $\Gamma$ as in \cref{fig: rolls}($c$), instead searching over $\Gamma$ only as needed to locate $\Gamma^*$ and $\Gamma^*_{loc}$. The rest of this section reports Nusselt number and Reynolds number scalings for the computed steady rolls, and the supplementary material provides tabulated data.

\subsection{Asymptotic heat transport}

\begin{figure}[t]
\centering
\includegraphics[width=0.65\textwidth]{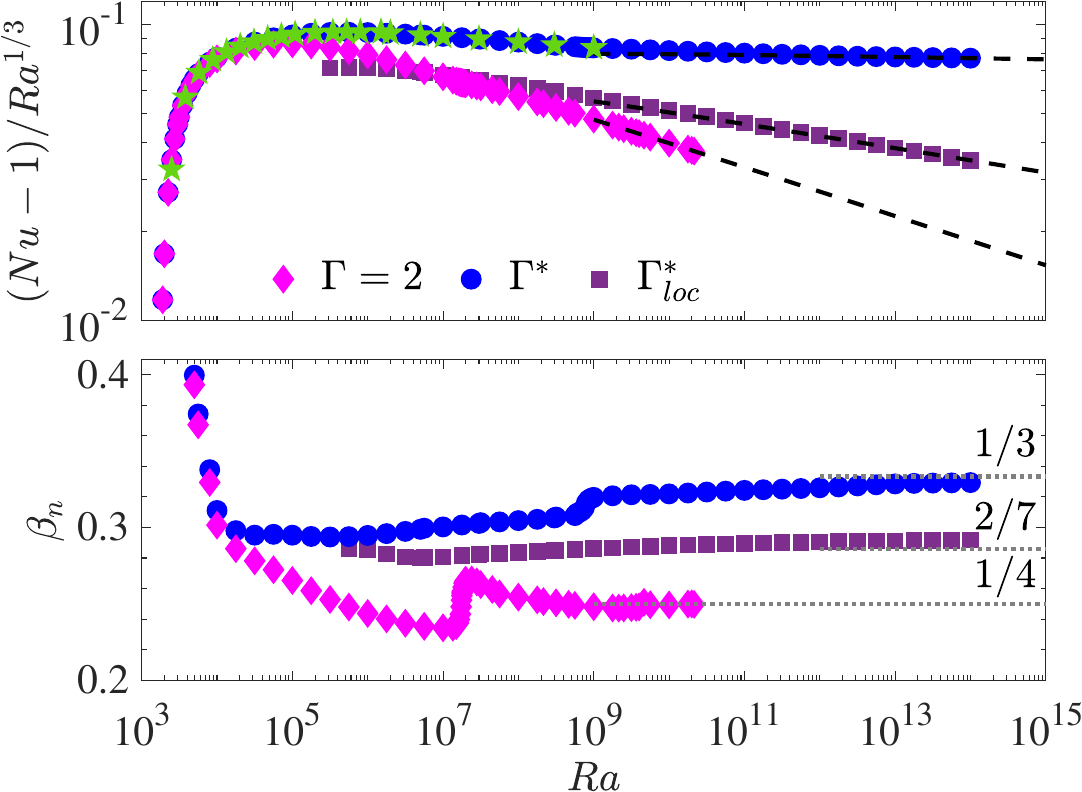}
\caption{\label{fig: Nu}Top: Compensated plot of $Nu-1$ vs.\ $Ra$ for steady rolls with $Pr=1$ and aspect ratios of $\Gamma=2$, $\Gamma^*$, and $\Gamma^*_{loc}$, where the $Ra$-dependent values $\Gamma^*$ and $\Gamma^*_{loc}$ are where $Nu(\Gamma)$ has global and local maxima, respectively (cf.\ figure~\ref{fig: rolls}). Values of $Nu$ at $\Gamma^*$ from \cite{Sondak2015} and \cite{Waleffe2020} are also shown ({\color{green}$\star$}). Scaling fits ($\dashedrule$) over the last decade of each data set yield exponents of $0.33$, $0.29$, and $0.25$.  Bottom: Finite difference approximations of the local scaling exponent $\beta_n = \frac{{\rm d}(\log Nu)}{{\rm d}(\log Ra)}$. Exponents of $1/3$, $2/7$, and $1/4$ are shown to guide the eye ({\color{mygray}$\dottedrule$}).}
\end{figure}

\Cref{fig: Nu} shows the dependence of $Nu$ on $Ra$ for steady rolls with aspect ratios $\Gamma=2$, $\Gamma^*$, and $\Gamma^*_{loc}$. In the top panel $Nu-1$ is compensated by $Ra^{1/3}$, so the horizontal line approached by rolls of the $Nu$-maximizing aspect ratios $\Gamma^*$ corresponds to classical $1/3$ scaling. The downward slopes of the data for aspect ratios 2 and $\Gamma^*_{loc}$ correspond to scaling exponents smaller than 1/3. Values of $Nu$ at $\Gamma^*$ computed previously for $Ra\le10^9$ \citep{Sondak2015,Waleffe2020} are shown in \cref{fig: Nu} also, and they agree with our computations very precisely---e.g., the $Ra=10^9$ data point agrees with our value of $Nu$ to within 0.0008\%.

The bottom panel of figure~\ref{fig: Nu} shows the $Ra$-dependent local scaling exponent $\beta_n = \frac{\rmd( \log Nu)}{\rmd (\log Ra)}$ of the $Nu$--$Ra$ relation for $\Gamma=2$, $\Gamma^*$, and $\Gamma^*_{loc}$. This quantity educes small variations not visible in the top panel.  In particular, for rolls of aspect ratios $\Gamma^*$, the exponent $\beta_n$ exhibits a small but rapid change just below $Ra = 10^9$, beyond which it smoothly approaches the classical $1/3$ exponent that appears to be the $Ra\to\infty$ asymptotic behavior.  This rapid change seems to coincide with the velocity becoming vertically uniform outside the boundary layers, as reflected in the streamlines of \cref{fig: rolls}($b$); further details of the rolls' structure will be reported elsewhere. Rolls with $\Gamma=2$ fixed undergo a similarly rapid change around $Ra \approx2 \times10^7$ and then approach $Nu \sim Ra^{1/4}$ scaling that appears to be asymptotic. Rolls of aspect ratio $\Gamma^*_{loc}$ show intermediate $Nu$ scaling whose best-fit exponent over the last decade of data is~0.29.

\begin{figure}[t]
\centering 
\includegraphics[width=0.65\textwidth]{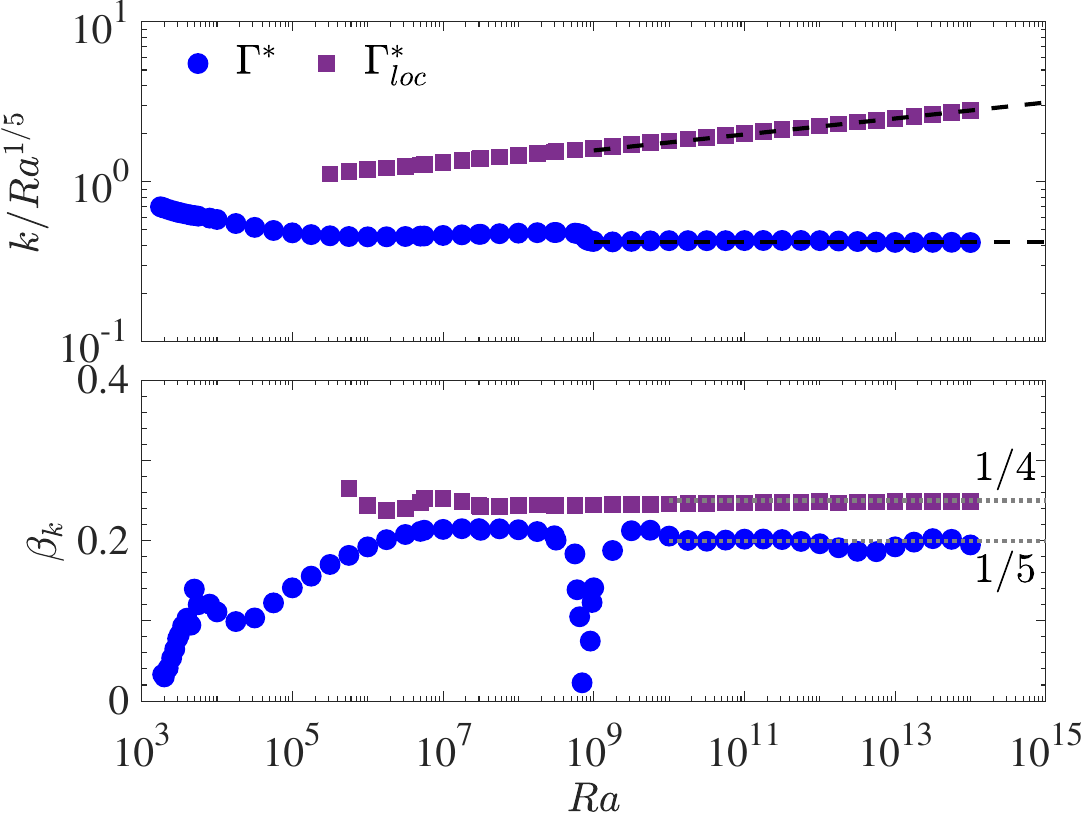}
\caption{\label{fig: k}Top: Compensated plot of the fundamental horizontal wavenumber $k=2\pi/\Gamma$ vs.\ $Ra$ for the aspect ratios $\Gamma^*$ and $\Gamma^*_{loc}$ that maximize $Nu(\Gamma)$ globally and locally, respectively, at $Pr=1$. Scaling fits ($\dashedrule$) to $2\pi/\Gamma^*$ over $Ra\in[10^{10},10^{14}]$ and $2\pi/\Gamma^*_{loc}$ over $Ra\in[10^{13},10^{14}]$ yield exponents of 0.20 and 0.25, respectively. Bottom: Finite difference approximations of the local exponent $\beta_k =  \frac{{\rm d}(\log k)}{{\rm d}(\log Ra)}$. The values $1/4$ and $1/5$ ({\color{mygray}$\dottedrule$}) agree with the scaling fit exponents to two digits.}
\end{figure}

The top panel of figure~\ref{fig: k} shows the $Ra$-dependence of the wavenumber $k=2\pi/\Gamma$ for $\Gamma^*$ and $\Gamma^*_{loc}$, compensated by $Ra^{1/5}$. The compensated wavenumbers for $\Gamma^*$ approach a horizontal line, suggesting that the $Nu$-maximizing rolls narrow according to the power law $\Gamma^*\sim Ra^{-1/5}$. This narrowing of $\Gamma^*$ is slow relative to the case of RBC in a porous medium, where $\Gamma^*\sim Ra^{-1/2}$ \citep{Wen2015JFM}.

The bottom panel of \cref{fig: k} shows the $Ra$-dependence of the local scaling exponent $\beta_k =  \frac{{\rm d}(\log k)}{{\rm d}(\log Ra)}$. For $k=2\pi/\Gamma^*$ the local scaling exponent remains close to $1/5$ after the transition around $Ra = 10^9$.  For $k=2\pi/\Gamma^*_{loc}$ the exponent seems to approach $1/4$, suggesting that $\Gamma^*_{loc}$ has the same $Ra^{-1/4}$ scaling as the narrowest marginally stable mode. Variations in $\beta_k$ beyond $Ra=10^{12}$ for $\Gamma^*$ are evident, but these might be due to numerical imprecision: $Nu$ depends very weakly on $\Gamma$ around the maximum of $Nu(\Gamma)$, as seen in \cref{fig: rolls}($c$), so the value of $\Gamma^*$ cannot be determined nearly as precisely as the value of $Nu(\Gamma^*)$.

\subsection{Asymptotic kinetic energy}

\begin{figure}[t]
\centering 
\includegraphics[width=0.65\textwidth]{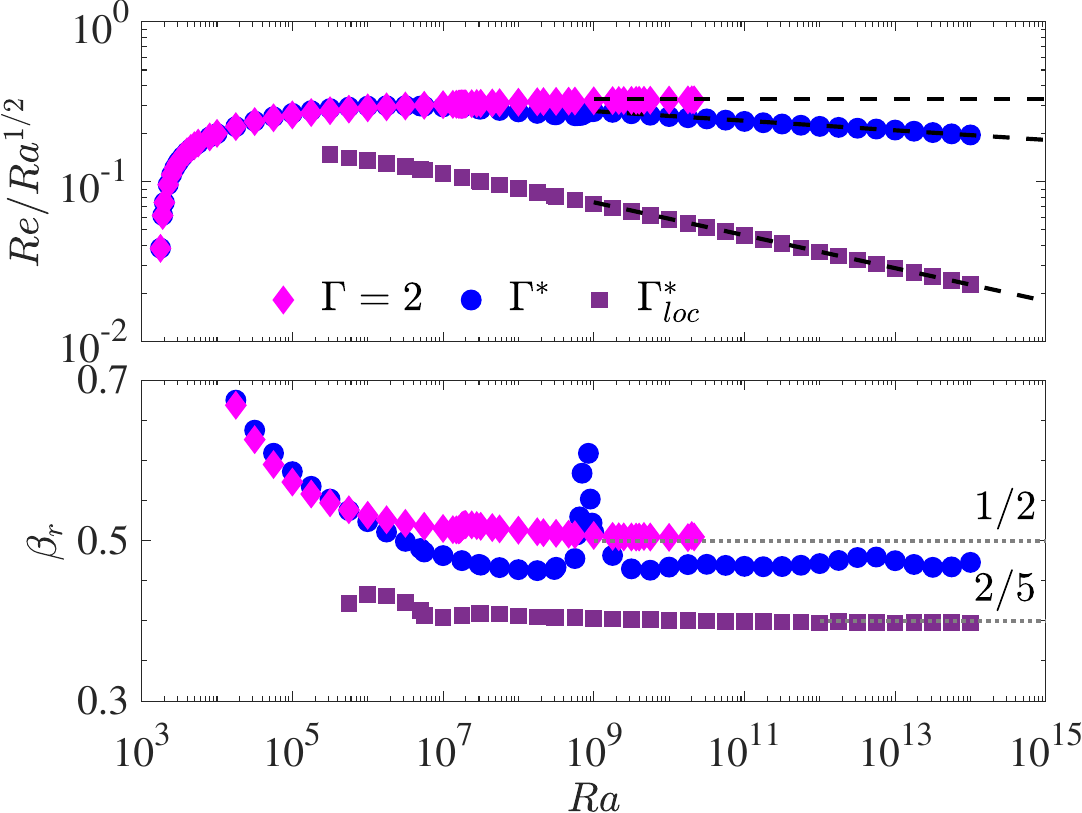} 
\caption{\label{fig: Re}Top: Compensated plot of $Re$ vs. $Ra$ for steady rolls with $Pr=1$ and aspect ratios $\Gamma=2$, $\Gamma^*$, and $\Gamma^*_{loc}$. Scaling fits yield $Re\sim Ra^{0.50}$ for $\Gamma=2$ and $Re\sim Ra^{0.40}$ for $\Gamma^*_{loc}$ over the last decade of each data set, and $Re\sim Ra^{0.47}$ for $\Gamma^*$ over $Ra\in[10^{10},10^{14}]$ ($\dashedrule$). Bottom: Finite difference approximations to the local exponent $\beta_r = \frac{{\rm d}( \log Re)}{{\rm d} (\log Ra)}$. Exponents of $1/2$ and $2/5$ are shown to guide the eye ({\color{mygray}$\dottedrule$}).}
\end{figure}

Another emergent quantity central to RBC is the bulk Reynolds number based on root-mean-squared velocity, which in terms of dimensionless solutions to \cref{eq: bouss} is
\beq
Re = \left( \frac{Ra}{Pr} \right)^{1/2} \langle | {\bf u}|^2 \rangle^{1/2}.
\label{eq: Re}
\eeq
\Cref{fig: Re} depicts the dependence of $Re$ on $Ra$ for the steady rolls of aspect ratios $\Gamma= 2$, $\Gamma^*$, and $\Gamma^*_{loc}$. The top panel shows $Re$ compensated by $Ra^{1/2}$ while the bottom panel shows the local scaling exponent $\beta_r =  \frac{{\rm d}( \log Re)}{{\rm d}( \log Ra)}$. Rolls with the fixed aspect ratio $\Gamma=2$ approach the asymptotic scaling $Re\sim Ra^{1/2}$ that corresponds to the root-mean-squared velocity being proportional to the free-fall velocity $U_f$. For rolls with $Nu$-maximizing aspect ratios $\Gamma^*$, the scaling fit over $Ra\in[10^{10},10^{14}]$ is $Re\sim Ra^{0.47}$, which is quite close to the $Re \sim Ra^{0.46}$ scaling observed in recent 3D direct numerical simulations up to $Ra=10^{15}$ at $Pr=1$ in a slender cylinder with a height 10 times its diameter \citep{Iyer2020}. For the $\Gamma^*_{loc}$ rolls the scaling exponent of $Re$ is indistinguishable from $2/5$.  The measured exponents (0.50, 0.47, 0.40) are unchanged if $Re$ is defined using the pointwise maximum velocity rather than using the root-mean-squared velocity as in \cref{eq: Re}.  All three aspect ratios result in smaller speeds than steady rolls between stress-free boundaries, where $Re\sim Ra^{2/3}$ for any fixed $Pr$ and $\Gamma$ \citep{Wen2020}.

\section{Comparison with turbulent convection}
\begin{figure}[t]
\centering 
\includegraphics[width=0.65\textwidth]{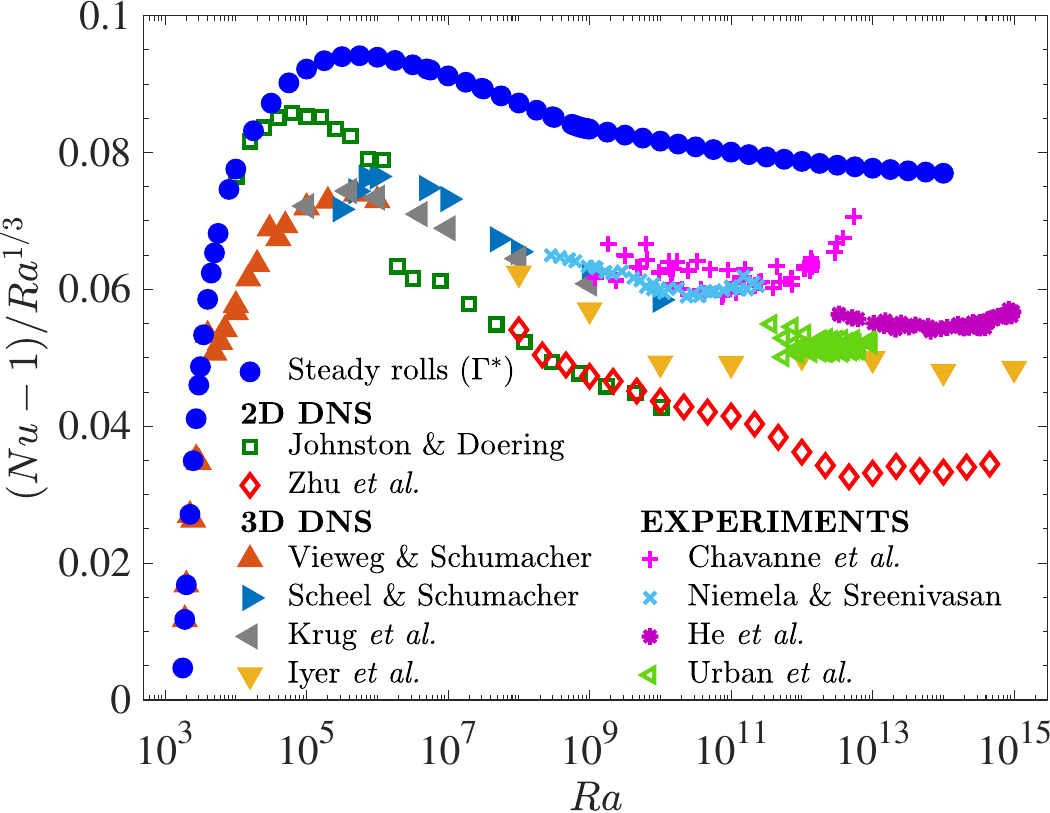}
\caption{\label{fig: turbulent}$Nu$ compensated by $Ra^{1/3}$ for steady rolls of $Nu$-maximizing aspect ratios $\Gamma^*$ at $\Pran=1$, along with $Nu$ from turbulent 2D and 3D DNS and experiments with estimated $\Pran \in [0.7,1.3]$. For horizontally periodic domains, 2D DNS with $(\Gamma,\Pran) = (2,1)$ were done by \citet{Johnston2009} and \citet{Zhu2018}, and 3D DNS with $\Gamma \ge 8$ and $\Pran = 1$ were done by \citet{Vieweg2020} and \citet{Krug2020}. For DNS in cylinders of diameter-to-height ratio $\Gamma_c$, \citet{Iyer2020} used $(\Gamma_c,\Pran)=(0.1,1)$ and \citet{Scheel2017} used $(\Gamma_c,\Pran)=(1,0.7)$. For laboratory experiments in cylinders, where the plotted data is truncated according to $\Pran\in[0.7,1.3]$, the domains and estimated $\Pran$ ranges are $\Gamma_c = 0.5$ and $\Pran\in[0.7,1.3]$ for \citet{Chavanne2001}, $\Gamma_c = 4$ and $\Pran\in[0.7,1.27]$ for \citet{Niemela2006},  $\Gamma_c = 0.5$ and $\Pran\in[0.79,0.86]$ for \citet{He2012PRL}, and $\Gamma_c = 1$ and $\Pran\in[0.95,1.17]$ for \citet{Urban2014}. Experiments used working fluids of low-temperature helium gas \citep{Chavanne2001, Niemela2006, Urban2014} or sulfur hexafluoride \citep{He2012PRL}.}
\end{figure}

To compare heat transport by steady rolls with that by turbulent thermal convection, we compiled Nusselt number data from high-$Ra$ DNS with $Pr=1$ or 0.7 and laboratory experiments where the estimated $Pr$ is between 0.7 and 1.3. \Cref{fig: turbulent} shows these $Nu$ values compensated by $Ra^{1/3}$, along with $Nu$ values of steady convection rolls at the $Nu$-maximizing aspect ratios $\Gamma^*$. Strikingly, heat transport by the $Nu$-maximizing 2D steady rolls is \emph{larger} than transport by turbulent convection in all cases.

The turbulent data shown in \cref{fig: turbulent}, as detailed in the figure caption, include DNS in horizontally periodic 2D and 3D domains, wherein 2D steady rolls solve the equations of motion, as well as 3D DNS and laboratory experiments in cylinders that do not admit 2D rolls. Values of $Nu$ for steady rolls with $\Gamma=2$ fixed are omitted from \cref{fig: turbulent} for clarity, but they lie below all turbulent values once $Ra$ approximately exceeds $2\times10^{9}$ (cf.\ \cref{fig: Nu}), and this gap would only widen at larger $Ra$ if their $Nu\sim Ra^{1/4}$ scaling persists. The laboratory data sets in \cref{fig: turbulent} have unavoidably varying $Pr$ values that can be hard to estimate, as well as non-Oberbeck--Boussinesq effects \citep{Urban2011, Urban2012, Urban2014}. The figure includes only a narrow range of estimated $Pr$ values in order to avoid significant non-Oberbeck--Boussinesq effects. When data over a wider range of estimated $Pr$ is included, a few data points from the experiments of \cite{Chavanne2001} lie above the $Nu(\Gamma^*)$ values of steady rolls, as shown in the supplementary material.

Our finding that steady rolls of $Nu$-maximizing aspect ratios apparently display classical $Nu\sim Ra^{1/3}$ asymptotic scaling does not ineluctably imply anything about turbulent convection. Taking a dynamical systems point of view, however, steady solutions admitted by the domain are fixed points of \cref{eq: bouss}, so they and their unstable manifolds are part of the global attractor.  Turbulent trajectories may linger near these fixed points and so inherit some quantitative features \citep{Kooloth2021}, as has been found for unstable coherent states in shear flows \citep{Nagata1990, Waleffe1998, Wedin2004, Gibson2008, SKGS2020, Graham2021}. Indeed, \cref{fig: turbulent} shows scaling similarities between steady and turbulent convection. Further exploration of the global attractor calls for study of 3D steady flows.  Recently computed `multi-scale' 3D steady states \citep{Motoki2021} give larger $Nu$ values than all 2D rolls at moderate $Ra$, but their scaling at large $Ra$ is unknown.  Simpler 3D steady convection patterns remain to be computed as well.   Analytically, it is an open challenge to construct approximations of 2D or 3D steady flows that are asymptotically accurate as $Ra\to\infty$, as has been done for 2D rolls between stress-free boundaries \citep{ChiniCox2009, Wen2020}.  Such constructions could be used to verify that $Nu\sim Ra^{1/3}$ is indeed the exact asymptotic scaling for the $Nu$-maximizing rolls we have computed, as well as to determine the precise $Re$--$Ra$ scaling relations for rolls of both $Nu$- and $Re$-maximizing aspect ratios.

More generally, \cref{fig: turbulent} highlights the absence of reproducible evidence for ultimate $Nu \sim Ra^{1/2}$ scaling, and it raises the intriguing possibility that \emph{steady} rolls with $Nu\sim Ra^{1/3}$ might transport more heat than turbulent convection as $Ra\to\infty$. We know of no counterexamples to this hypothesis, including in the case of stress-free boundaries \citep{Wen2020}. Heat transport by solutions of \cref{eq: bouss} with no-slip isothermal boundaries has been mathematically proved to be limited by $Nu\le\cO(Ra^{1/2})$ \citep{H63,DC1996}, but it remains unknown whether any solutions attain the ultimate scaling of this upper bound.  One avenue for pursuing a stronger mathematical statement is to study two conjectures suggested by our computations: that steady convection maximizes $Nu$ among all solutions of \cref{eq: bouss} regardless of their stability or time-dependence, and that steady solutions of \cref{eq: bouss} are subject to an upper bound of the form $Nu\le\cO(Ra^{1/3})$. Therefore, although numerically computed flows can never determine $Ra\to\infty$ scaling definitively, our results suggest a new mathematical approach that may be able to finally resolve the question of asymptotic $Nu$ scaling in turbulent convection.

\vspace{-0.15in}
\section*{Acknowledgements}
After this manuscript was written our senior author, Charles Doering, passed away too soon. Beyond his many contributions to the present study, we are forever indebted to him for his mentorship, to say nothing of his many lasting contributions to the field of fluid dynamics. He will be deeply missed by us and many others. We also want to acknowledge helpful discussions about the present work with L.M.\ Smith, D.\ Sondak, and F.\ Waleffe. This work was supported by US National Science Foundation awards (DMS-1515161, DMS-1813003), Canadian NSERC Discovery Grants Program awards (RGPIN-2018-04263, RGPAS-2018-522657, DGECR-2018-00371), and computational resources provided by Advanced Research Computing at the University of Michigan.

\vspace{-0.15in}
\section*{Declaration of interests}
The authors report no conflict of interest.
\vspace{-0.15in}

\newpage

\section*{Supplementary Material}
\addcontentsline{toc}{section}{\protect\numberline{}Supplemental Material}

\subsection*{Numerical solutions}

Tables~\ref{tab:Gamma2} to \ref{tab:GammaLocal} give the $Nu$, $k$, and $Re$ values for numerical solutions with $Pr=1$ and $\Gamma = 2$, $\Gamma^*$, and $\Gamma^*_{loc}$, respectively.\vspace{-12pt}

\begin{longtable}{cccccc}
\caption{Details for numerical solutions with $Pr=1$ and $\Gamma=2$, including the resolution of Fourier modes ($N_x$) and Chebyshev collocation points ($N_z$).} \\
{\label{tab:Gamma2}}
 \quad $Ra$	\quad  & \quad $Pr$ \quad & \quad $k=2\pi/\Gamma$ \quad &  \quad\quad $N_x \times N_z$ \quad\quad & \quad\quad\quad $Nu$ \quad\quad\quad & \quad\quad\quad $Re$	\quad\quad\quad\\
$10^{13/4}$	&	1	&	$\pi$	&	128 $\times$ 65	&	1.056590	&	1.617336\\
$1.9\times10^3$	&	1	&	$\pi$	&	128 $\times$ 65	&	1.145807	&	2.682155\\
$2\times10^3$	&	1	&	$\pi$	&	128 $\times$ 65	&	1.212037	&	3.317190\\
$2.25\times10^3$	&	1	&	$\pi$	&	128 $\times$ 65	&	1.355410	&	4.550975\\
$2.5\times10^3$	&	1	&	$\pi$	&	128 $\times$ 65	&	1.474455	&	5.537770\\
$2.75\times10^3$	&	1	&	$\pi$	&	128 $\times$ 65	&	1.575599	&	6.391812\\
$3\times10^3$	&	1	&	$\pi$	&	128 $\times$ 65	&	1.663162	&	7.159844\\
$10^{14/4}$	&	1	&	$\pi$	&	128 $\times$ 65	&	1.714193	&	7.624400\\
$3.5\times10^3$	&	1	&	$\pi$	&	128 $\times$ 65	&	1.808754	&	8.526064\\
$4\times10^3$	&	1	&	$\pi$	&	128 $\times$ 65	&	1.926775	&	9.740578\\
$4.5\times10^3$	&	1	&	$\pi$	&	128 $\times$ 65	&	2.025985	&	10.85141\\
$5\times10^3$	&	1	&	$\pi$	&	128 $\times$ 65	&	2.111714	&	11.88534\\
$10^{15/4}$	&	1	&	$\pi$	&	128 $\times$ 65	&	2.204811	&	13.09152\\
$8\times10^3$	&	1	&	$\pi$	&	128 $\times$ 65	&	2.476330	&	17.05494\\
$10^4$	&	1	&	$\pi$	&	128 $\times$ 65	&	2.648664	&	20.07400\\
$10^{17/4}$	&	1	&	$\pi$	&	128 $\times$ 65	&	3.122843	&	29.50047\\
$10^{18/4}$	&	1	&	$\pi$	&	128 $\times$ 65	&	3.665041	&	42.29585\\
$10^{19/4}$	&	1	&	$\pi$	&	128 $\times$ 65	&	4.287042	&	59.56858\\
$10^{5}$	&	1	&	$\pi$	&	128 $\times$ 65	&	4.994322	&	82.84462\\
$10^{21/4}$	&	1	&	$\pi$	&	128 $\times$ 65	&	5.795869	&	114.2355\\
$10^{22/4}$	&	1	&	$\pi$	&	128 $\times$ 97	&	6.703915	&	156.5252\\
$10^{23/4}$	&	1	&	$\pi$	&	128 $\times$ 97	&	7.732236	&	213.4031\\
$10^{6}$	&	1	&	$\pi$	&	128 $\times$ 97	&	8.896615	&	289.7982\\
$10^{25/4}$	&	1	&	$\pi$	&	256 $\times$ 97	&	10.21546	&	392.2837\\
$10^{26/4}$	&	1	&	$\pi$	&	256 $\times$ 129	&	11.71065	&	529.6372\\
$10^{27/4}$	&	1	&	$\pi$	&	256 $\times$ 129	&	13.40898	&	713.6005\\
$10^{7}$	&	1	&	$\pi$	&	512 $\times$ 129	&	15.34493	&	959.9367\\
$1.35\times10^{7}$	&	1	&	$\pi$	&	512 $\times$ 129	&	16.46456	&	1119.932\\
$1.5\times10^{7}$	&	1	&	$\pi$	&	512 $\times$ 129	&	16.87881	&	1182.172\\
$1.6\times10^{7}$	&	1	&	$\pi$	&	512 $\times$ 193	&	17.13944	&	1222.005\\
$1.65\times10^{7}$	&	1	&	$\pi$	&	512 $\times$ 193	&	17.26636	&	1241.484\\
$1.7\times10^{7}$	&	1	&	$\pi$	&	512 $\times$ 193	&	17.39216	&	1260.709\\
$1.736\times10^{7}$	&	1	&	$\pi$	&	512 $\times$ 193	&	17.48282	&	1274.414\\
$1.76\times10^{7}$	&	1	&	$\pi$	&	512 $\times$ 193	&	17.54351	&	1283.493\\
$10^{29/4}$	&	1	&	$\pi$	&	512 $\times$ 193	&	17.58987	&	1290.377\\
$1.786\times10^{7}$	&	1	&	$\pi$	&	512 $\times$ 193	&	17.60946	&	1293.277\\
$1.8\times10^{7}$	&	1	&	$\pi$	&	512 $\times$ 193	&	17.64500	&	1298.522\\
$1.85\times10^{7}$	&	1	&	$\pi$	&	512 $\times$ 193	&	17.77160	&	1317.122\\
$1.9\times10^{7}$	&	1	&	$\pi$	&	512 $\times$ 193	&	17.89693	&	1335.501\\
$1.95\times10^{7}$	&	1	&	$\pi$	&	512 $\times$ 193	&	18.02044	&	1353.657\\
$2\times10^{7}$	&	1	&	$\pi$	&	512 $\times$ 193	&	18.14193	&	1371.593\\
$2.4\times10^{7}$	&	1	&	$\pi$	&	512 $\times$ 193	&	19.04221	&	1507.890\\
$2.8\times10^{7}$	&	1	&	$\pi$	&	512 $\times$ 193	&	19.83413	&	1633.418\\
$10^{30/4}$	&	1	&	$\pi$	&	512 $\times$ 193	&	20.47798	&	1739.647\\
$4.5\times10^{7}$	&	1	&	$\pi$	&	512 $\times$ 193	&	22.44036	&	2087.182\\
$10^{31/4}$	&	1	&	$\pi$	&	512 $\times$ 193	&	23.76002	&	2340.822\\
$10^{8}$	&	1	&	$\pi$	&	768 $\times$ 257	&	27.50669	&	3144.931\\
$10^{33/4}$	&	1	&	$\pi$	&	768 $\times$ 257	&	31.81154	&	4220.616\\
$2.15\times10^{8}$	&	1	&	$\pi$	&	768 $\times$ 257	&	33.36657	&	4649.541\\
$10^{34/4}$	&	1	&	$\pi$	&	768 $\times$ 257	&	36.75427	&	5658.648\\
$4.64\times10^{8}$	&	1	&	$\pi$	&	896 $\times$ 321	&	40.44652	&	6875.704\\
$10^{35/4}$	&	1	&	$\pi$	&	896 $\times$ 321	&	42.42917	&	7580.057\\
$10^{9}$	&	1	&	$\pi$	&	1024 $\times$ 321	&	48.94284	&	10145.79\\
$10^{37/4}$	&	1	&	$\pi$	&	1024 $\times$ 321	&	56.42926	&	13571.10\\
$2.15\times10^{9}$	&	1	&	$\pi$	&	1024 $\times$ 321	&	59.13988	&	14935.81\\
$2.5\times10^{9}$	&	1	&	$\pi$	&	1024 $\times$ 321	&	61.38714	&	16116.94\\
$10^{38/4}$	&	1	&	$\pi$	&	1024 $\times$ 321	&	65.06338	&	18145.61\\
$10^{153/16}$	&	1	&	$\pi$	&	1024 $\times$ 321	&	67.42466	&	19511.79\\
$4\times10^{9}$	&	1	&	$\pi$	&	1024 $\times$ 321	&	68.96487	&	20429.24\\
$4.64\times10^{9}$	&	1	&	$\pi$	&	1024 $\times$ 321	&	71.58241	&	22019.93\\
$10^{39/4}$	&	1	&	$\pi$	&	1024 $\times$ 321	&	75.09725	&	24262.22\\
$10^{10}$	&	1	&	$\pi$	&	1024 $\times$ 321	&	86.68318	&	32430.06\\
$10^{41/4}$	&	1	&	$\pi$	&	1024 $\times$ 321	&	100.0909	&	43339.02\\
$2\times10^{10}$	&	1	&	$\pi$	&	1024 $\times$ 321	&	103.0738	&	45995.60\\
$2.15\times10^{10}$ &	1	&	$\pi$	&	1024 $\times$ 321	&	104.9517	&	47705.08
\end{longtable}
\vspace{-12pt}

\begin{longtable}{cccccc}
\caption{Details for numerical solutions with $Pr=1$ and the aspect ratios $\Gamma^*$ that globally maximize $Nu(\Gamma)$, including the resolution of Fourier modes ($N_x$) and Chebyshev collocation points ($N_z$).} \\
{\label{tab:GammaGlobal}}
\quad $Ra$ \quad &	\quad $Pr$ \quad & \quad $k=2\pi/\Gamma^*$ \quad		&	\quad\quad $N_x \times N_z$	\quad\quad  &	\quad\quad\quad		$Nu$	\quad\quad\quad		&	\quad\quad\quad  $Re$	\quad\quad\quad\\
$10^{13/4}$	&	1	&	3.116683	&	128 $\times$ 65	&	1.056697	&	1.619793\\
$1.9\times10^{3}$	&	1	&	3.123537	&	128 $\times$ 65	&	1.145870	&	2.683859\\
$2\times10^{3}$	&	1	&	3.128360	&	128 $\times$ 65	&	1.212070	&	3.318462\\
$2.25\times10^{3}$	&	1	&	3.143491	&	128 $\times$ 65	&	1.355411	&	4.550777\\
$2.5\times10^{3}$	&	1	&	3.161280	&	128 $\times$ 65	&	1.474516	&	5.535574\\
$2.75\times10^{3}$	&	1	&	3.180831	&	128 $\times$ 65	&	1.575828	&	6.387203\\
$3\times10^{3}$	&	1	&	3.202416	&	128 $\times$ 65	&	1.663668	&	7.152347\\
$10^{14/4}$	&	1	&	3.216383	&	128 $\times$ 65	&	1.714937	&	7.614958\\
$3.5\times10^{3}$	&	1	&	3.247094	&	128 $\times$ 65	&	1.810118	&	8.512058\\
$4\times10^{3}$	&	1	&	3.292192	&	128 $\times$ 65	&	1.929322	&	9.719380\\
$4.5\times10^{3}$	&	1	&	3.329096	&	128 $\times$ 65	&	2.029942	&	10.82473\\
$5\times10^{3}$	&	1	&	3.378413	&	128 $\times$ 65	&	2.117243	&	11.84892\\
$10^{15/4}$	&	1	&	3.426419	&	128 $\times$ 65	&	2.212421	&	13.04623\\
$8\times10^{3}$	&	1	&	3.575467	&	128 $\times$ 65	&	2.492199	&	17.05494\\
$10^{4}$	&	1	&	3.665236	&	128 $\times$ 65	&	2.671348	&	19.99537\\
$10^{17/4}$	&	1	&	3.880392	&	128 $\times$ 65	&	3.171063	&	29.49222\\
$10^{18/4}$	&	1	&	4.118841	&	128 $\times$ 65	&	3.757873	&	42.57254\\
$10^{19/4}$	&	1	&	4.419469	&	128 $\times$ 89	&	4.454688	&	60.44520\\
$10^{5}$	&	1	&	4.793529	&	128 $\times$ 89	&	5.278963	&	84.69567\\
$10^{21/4}$	&	1	&	5.242992	&	128 $\times$ 129	&	6.252782	&	117.4323\\
$10^{22/4}$	&	1	&	5.782949	&	128 $\times$ 129	&	7.404680	&	161.3367\\
$10^{23/4}$	&	1	&	6.420133	&	128 $\times$ 129	&	8.769978	&	219.8304\\
$10^{6}$	&	1	&	7.171170	&	128 $\times$ 129	&	10.39171	&	297.1691\\
$10^{25/4}$	&	1	&	8.051634	&	128 $\times$ 129	&	12.32188	&	398.7272\\
$10^{26/4}$	&	1	&	9.073939	&	128 $\times$ 129	&	14.62274	&	531.4357\\
$5\times10^{6}$	&	1	&	9.997275	&	192 $\times$ 257	&	16.76769	&	665.2339\\
$10^{27/4}$	&	1	&	10.25059	&	192 $\times$ 257	&	17.36827	&	704.3108\\
$10^{7}$	&	1	&	11.59439	&	192 $\times$ 257	&	20.64616	&	929.1624\\
$10^{29/4}$	&	1	&	13.12083	&	192 $\times$ 257	&	24.56044	&	1221.405\\
$3\times10^{7}$	&	1	&	14.68186	&	256 $\times$ 257	&	28.77198	&	1562.064\\
$10^{30/4}$	&	1	&	14.84741	&	256 $\times$ 257	&	29.23512	&	1601.174\\
$10^{31/4}$	&	1	&	16.80072	&	256 $\times$ 257	&	34.81847	&	2094.287\\
$10^{8}$	&	1	&	18.99815	&	256 $\times$ 321	&	41.48855	&	2735.227\\
$10^{33/4}$	&	1	&	21.45545	&	256 $\times$ 321	&	49.46027	&	3569.756\\
$3\times10^{8}$	&	1	&	23.89666	&	256 $\times$ 321	&	58.05030	&	4550.127\\
$10^{34/4}$	&	1	&	24.15059	&	256 $\times$ 321	&	58.99612	&	4663.393\\
$10^{35/4}$	&	1	&	26.84021	&	256 $\times$ 321	&	70.43089	&	6139.224\\
$6\times10^{8}$	&	1	&	27.08270	&	256 $\times$ 321	&	71.85714	&	6344.418\\
$6.5\times10^{8}$	&	1	&	27.31152	&	256 $\times$ 321	&	73.66207	&	6619.142\\
$7\times10^{8}$	&	1	&	27.35757	&	256 $\times$ 321	&	75.37932	&	6911.933\\
$7.5\times10^{8}$	&	1	&	26.83143	&	256 $\times$ 321	&	77.02411	&	7294.891\\
$8\times10^{8}$	&	1	&	26.28179	&	256 $\times$ 321	&	78.61210	&	7683.031\\
$8.5\times10^{8}$	&	1	&	26.25586	&	256 $\times$ 321	&	80.14209	&	7971.925\\
$9\times10^{8}$	&	1	&	26.36825	&	256 $\times$ 321	&	81.61573	&	8227.378\\
$9.5\times10^{8}$	&	1	&	26.54403	&	256 $\times$ 321	&	83.03697	&	8462.830\\
$10^{9}$	&	1	&	26.73696	&	256 $\times$ 321	&	84.40976	&	8687.394\\
$10^{37/4}$	&	1	&	29.78702	&	512 $\times$ 449	&	101.5246	&	11462.12\\
$10^{38/4}$	&	1	&	33.65968	&	512 $\times$ 449	&	122.1559	&	14978.49\\
$10^{39/4}$	&	1	&	38.04901	&	512 $\times$ 449	&	146.9986	&	19554.08\\
$10^{10}$	&	1	&	42.83017	&	512 $\times$ 449	&	176.9293	&	25585.09\\
$10^{41/4}$	&	1	&	48.06231	&	512 $\times$ 449	&	213.0247	&	33536.66\\
$10^{42/4}$	&	1	&	53.90331	&	512 $\times$ 449	&	256.5802	&	43967.29\\
$10^{43/4}$	&	1	&	60.50367	&	512 $\times$ 513	&	309.1454	&	57597.49\\
$10^{11}$	&	1	&	67.95755	&	512 $\times$ 513	&	372.5844	&	75402.24\\
$10^{45/4}$	&	1	&	76.33729	&	512 $\times$ 769	&	449.1508	&	98685.64\\
$10^{46/4}$	&	1	&	85.71701	&	512 $\times$ 897	&	541.5753	&	129175.9\\
$10^{47/4}$	&	1	&	96.12138	&	512 $\times$ 897	&	653.1727	&	169237.3\\
$10^{12}$	&	1	&	107.6085	&	512 $\times$ 897	&	787.9764	&	221995.2\\
$10^{49/4}$	&	1	&	120.1234	&	512 $\times$ 897	&	950.9070	&	291852.5\\
$10^{50/4}$	&	1	&	133.7508	&	512 $\times$ 897	&	1147.971	&	384498.9\\
$10^{51/4}$	&	1	&	148.8836	&	512 $\times$ 1025	&	1386.450	&	506738.9\\
$10^{13}$	&	1	&	166.3042	&	512 $\times$ 1025	&	1675.036	&	666086.2\\
$10^{53/4}$	&	1	&	186.3974	&	512 $\times$ 1025	&	2024.094	&	873176.4\\
$10^{54/4}$	&	1	&	209.4395	&	512 $\times$ 1281	&	2446.172	&	1142290\\
$10^{55/4}$	&	1	&	235.2120	&	512 $\times$ 1537	&	2956.470	&	1494811\\
$10^{14}$		&	1	&	263.0987	&	512 $\times$ 1793	&	3573.640	&	1962459
\end{longtable}
\vspace{0pt}

\begin{longtable}{cccccc}
\caption{Details for numerical solutions with $Pr=1$ and the aspect ratios $\Gamma^*_{loc}$ that locally maximize $Nu(\Gamma)$, including the resolution of Fourier modes ($N_x$) and Chebyshev collocation points ($N_z$).} \\
{\label{tab:GammaLocal}}
\quad $Ra$ \quad &	\quad $Pr$ \quad &	\quad  $k=2\pi/\Gamma^*_{loc}$ \quad &	\quad\quad $N_x \times N_z$	\quad\quad &	\quad\quad\quad $Nu$ \quad\quad\quad &	\quad\quad\quad		$Re$	\quad\quad\quad\\
$10^{22/4}$	&	1	&	14.09456	&	96 $\times$ 129	&	5.864201	&	82.92705\\
$10^{23/4}$	&	1	&	16.41959	&	96 $\times$ 129	&	6.914669	&	105.6969\\
$10^{6}$	&	1	&	18.89401	&	96 $\times$ 129	&	8.148261	&	135.6083\\
$10^{25/4}$	&	1	&	21.66773	&	96 $\times$ 129	&	9.587445	&	173.8004\\
$10^{26/4}$	&	1	&	24.88046	&	96 $\times$ 129	&	11.26803	&	221.7384\\
$5\times10^{6}$	&	1	&	27.86575	&	96 $\times$ 129	&	12.81012	&	267.8677\\
$10^{27/4}$	&	1	&	28.70377	&	96 $\times$ 129	&	13.23878	&	280.9653\\
$10^{7}$	&	1	&	33.19466	&	96 $\times$ 129	&	15.56038	&	354.6158\\
$10^{29/4}$	&	1	&	38.29704	&	96 $\times$ 129	&	18.29931	&	448.0501\\
$3\times10^{7}$	&	1	&	43.50700	&	96 $\times$ 193	&	21.21050	&	554.8653\\
$10^{30/4}$	&	1	&	44.06680	&	96 $\times$ 193	&	21.52859	&	566.9565\\
$10^{31/4}$	&	1	&	50.67686	&	96 $\times$ 193	&	25.33533	&	717.1889\\
$10^{8}$	&	1	&	58.31124	&	96 $\times$ 193	&	29.82540	&	906.1212\\
$10^{33/4}$	&	1	&	67.11615	&	128 $\times$ 321	&	35.12519	&	1143.903\\
$3\times10^{8}$	&	1	&	76.25123	&	128 $\times$ 321	&	40.76618	&	1413.329\\
$10^{34/4}$	&	1	&	77.23740	&	128 $\times$ 321	&	41.38311	&	1443.747\\
$10^{35/4}$	&	1	&	88.88239	&	128 $\times$ 321	&	48.77387	&	1821.661\\
$10^{9}$	&	1	&	102.3071	&	128 $\times$ 321	&	57.50461	&	2297.406\\
$10^{37/4}$	&	1	&	117.7822	&	128 $\times$ 449	&	67.82060	&	2896.298\\
$10^{38/4}$	&	1	&	135.6225	&	128 $\times$ 449	&	80.01178	&	3650.131\\
$10^{39/4}$	&	1	&	156.1963	&	128 $\times$ 449	&	94.42106	&	4598.740\\
$10^{10}$	&	1	&	179.9312	&	128 $\times$ 449	&	111.4542	&	5792.111\\
$10^{41/4}$	&	1	&	207.3119	&	128 $\times$ 449	&	131.5910	&	7293.324\\
$10^{42/4}$	&	1	&	238.9044	&	128 $\times$ 449	&	155.3995	&	9181.462\\
$10^{43/4}$	&	1	&	275.3612	&	128 $\times$ 449	&	183.5515	&	11555.93\\
$10^{11}$	&	1	&	317.4310	&	128 $\times$ 449	&	216.8420	&	14541.83\\
$10^{45/4}$	&	1	&	365.9813	&	128 $\times$ 641	&	256.2116	&	18296.30\\
$10^{46/4}$	&	1	&	422.0132	&	128 $\times$ 641	&	302.7737	&	23016.84\\
$10^{47/4}$	&	1	&	486.6804	&	128 $\times$ 641	&	357.8457	&	28951.78\\
$10^{12}$	&	1	&	561.6657	&	128 $\times$ 641	&	422.9866	&	36390.82\\
$10^{49/4}$	&	1	&	647.4534	&	128 $\times$ 641	&	500.0423	&	45793.62\\
$10^{50/4}$	&	1	&	746.8566	&	128 $\times$ 641	&	591.1965	&	57587.15\\
$10^{51/4}$	&	1	&	861.4431	&	128 $\times$ 769	&	699.0342	&	72424.84\\
$10^{13}$	&	1	&	994.2189	&	128 $\times$ 769	&	826.6155	&	91031.10\\
$10^{53/4}$	&	1	&	1147.050	&	128 $\times$ 769	&	977.5620	&	114458.7\\
$10^{54/4}$	&	1	&	1323.461	&	128 $\times$ 1025	&	1156.161	&	143907.6\\
$10^{55/4}$	&	1	&	1527.004	&	128 $\times$ 1025	&	1367.486	&	180935.2\\
$10^{14}$	&	1	&	1762.395	&	128 $\times$ 1025	&	1617.546	&	227422.7
\end{longtable}

\begin{figure}[h]
\centering
\includegraphics[width=0.65\textwidth]{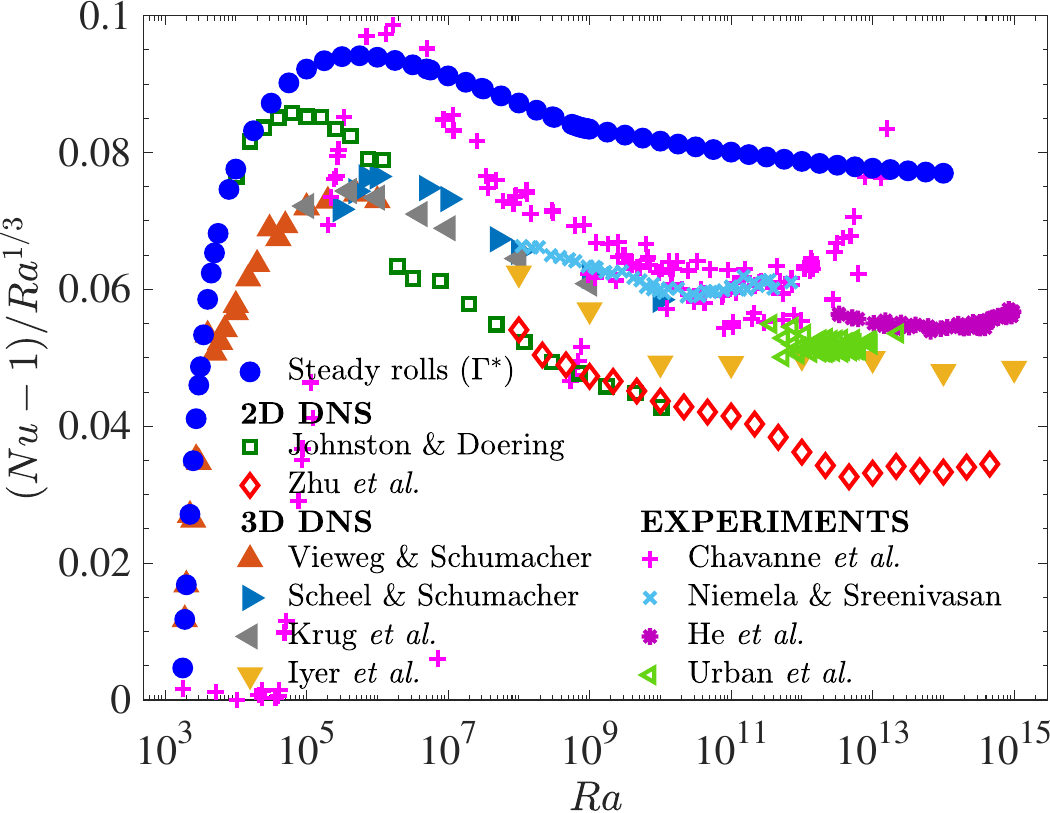}
\caption*{Figure 1S.  $Nu$ compensated by $Ra^{1/3}$ for steady rolls of $Nu$-maximizing aspect ratios $\Gamma^*$ at $\Pran=1$, along with $Nu$ from turbulent 2D and 3D DNS and experiments with estimated $\Pran \in [0.6,2]$. For horizontally periodic domains, 2D DNS with $(\Gamma,\Pran) = (2,1)$ were done by \citet{Johnston2009} and \citet{Zhu2018}, and 3D DNS with $\Gamma \ge 8$ and $\Pran = 1$ were done by \citet{Vieweg2020} and \citet{Krug2020}. For DNS in cylinders of diameter-to-height ratio $\Gamma_c$, \citet{Iyer2020} used $(\Gamma_c,\Pran)=(0.1,1)$ and \citet{Scheel2017} used $(\Gamma_c,\Pran)=(1,0.7)$. For laboratory experiments in cylinders, where the plotted data is truncated according to $\Pran\in[0.5,2]$, the domains and estimated $\Pran$ ranges are $\Gamma_c = 0.5$ and $\Pran\in[0.6,2]$ for \citet{Chavanne2001}, $\Gamma_c = 4$ and $\Pran\in[0.69,1.84]$ for \citet{Niemela2006},  $\Gamma_c = 0.5$ and $\Pran\in[0.79,0.86]$ for \citet{He2012PRL}, and $\Gamma_c = 1$ and $\Pran\in[0.95,1.5]$ for \citet{Urban2014}. Experiments used working fluids of low-temperature helium gas \citep{Chavanne2001, Niemela2006, Urban2014} or sulfur hexafluoride \citep{He2012PRL}.}
\end{figure}

\subsection*{Comparison with turbulent convection} 

Figure~1S is nearly identical to figure~5 in the main text, comparing heat transport by $Nu$-maximizing steady rolls with transport by turbulent convection, except that more experimental data with Prandtl numbers further from 1 are included.  In figure~1S the criterion for inclusion is an estimated Prandtl number of $Pr\in[0.5,2]$ rather than the range $\Pran\in[0.7,1.3]$ in figure~5 of the main text. (In fact all of the estimated $\Pran$ are at least 0.6, so $\Pran\in[0.6,2]$ in figure~1S.) The working fluids in the experiments---gaseous helium or sulfur hexaflouride---are used near their critical points, leading to coupling and sensitive variation of material parameters that can be difficult to estimate.  Faster variation of $Pr$ with $Ra$ is associated with increasing non-Oberbeck--Boussinesq effects as well; see \cite{Urban2011, Urban2012, Urban2014} for a discussion of experimental challenges. Data in figure~5 is truncated using the narrower range $Pr\in[0.7,1.3]$ mainly to reduce non-Oberbeck--Boussinesq effects---we expect $\Pran$ alone to have a more modest effect, even over the wider range $[0.5,2]$.

\newpage

\bibliographystyle{jfm}
\bibliography{no_slip}

\end{document}